\documentclass[11pt]{article}

\usepackage[T1]{fontenc} 
\usepackage[latin1]{inputenc}
\usepackage{amsmath}
\usepackage{amsfonts}
\usepackage{amssymb}
\usepackage{alltt}        
\usepackage{verbatim}
\usepackage{a4wide}
\usepackage{xspace}
\usepackage{booktabs}
\usepackage{url}
\usepackage{color}
\usepackage{listings}

\usepackage{enumitem}

\usepackage[protrusion=true,expansion=true]{microtype}  

\usepackage{graphicx, subfig}
\usepackage{cite}

\newcommand{\GeV}{{\rm\ GeV}}

\newcommand{\TeV}{{\rm\ TeV}}

\definecolor{comment}{rgb}{0,0.3,0}
\definecolor{identifier}{rgb}{0.0,0,0.3}

\newcommand{\ttt}[1]{{\small\texttt{#1}}}
\newcommand{\template}{{\sc TemplateTagger}\xspace}

\begin{document}

\begin{center}

{\large { \sc TemplateTagger v1.0.0} \\
A Template Matching Tool for Jet Substructure}
\end{center}

\begin{center}
 {Mihailo Backovi\'{c}},  {Jos\'{e} Juknevich}\\

\vskip 8pt

    {\it \small Department of Particle Physics and Astrophysics \\
   Weizmann Institute of Science, Rehovot 76100, Israel}\\

\vspace*{0.853cm}\end{center}

\begin{abstract}

 {\sc TemplateTagger} is a {\sc C++}  package for jet substructure analysis with Template Overlap Method. 
 The code operates with arbitrary models within fixed-order perturbation theory and arbitrary kinematics. Specialized template generation classes allow the user to implement any model for a decay of a boosted heavy object. In addition to Template Overlap, the code provides ability to calculate other template shape and energy flow observables. We describe in detail the structure of the package, as well as provide examples of its usage. 
\end{abstract}
\newpage

\tableofcontents

\section{Introduction}

Algorithms for tagging of boosted objects necessarily exploit observables sensitive to parton shower history, including color flow and hadronization. Most jet sub-structure methods can be characterized as jet de-clustering and re-clustering algorithms, with a common feature that they perform the analysis on the entire jet, after showering and hadronization. (see Refs. \cite{Ellis:2007ib, Abdesselam:2010pt, Altheimer:2012mn,  Salam:2009jx, Nath:2010zj,Plehn:2011tg, Thaler:2011gf, Chen:2011ah, Han:2011ab, Feige:2012vc, Plehn:2009rk,Plehn:2011sj, Plehn:2011tf, Thaler:2010tr, Cui:2010km, Hook:2011cq, Soper:2012pb, Kaplan:2008ie,Gallicchio:2010sw} for a review).

Tracking the parton shower history from physical final states to the hard-parton subprocesses often becomes rather involved as the number of QCD emissions is typically very large. Details of hadronization only further complicate the analysis. The  Template Overlap Method \cite{Almeida:2011aa, Almeida:2011ud,Almeida:2008tp,Almeida:2010pa}, aims to bridge the gap between energy flow of observed jets
and partonic configurations calculated at fixed order perturbation theory. The method allows for subjet identification in an infrared-safe way, by providing a mapping between energy-unweighted variables and the template that defines the energy flow distribution. 

The primary intended use of the \template package is the analysis of jet substructure in High Energy Physics collider data. \template allows the user to design a custom boosted jet analysis for a variety of scenarios by using the same basic three-stage approach. First, the user generates sets (or ``catalogs'') of templates by scanning over a phase space of the parton decay daughters of a massive particle of mass $M$ and transverse momentum $p_T$. Second, the  \template code performs template matching on an event-by-event basis whereby candidate signal jets are located in the $\eta,\phi, p_T$ space. Finally, the events are tagged using best matched templates (peak templates) as approximate subjet locations. The overlap approach has several important features:

\begin{itemize}
\item \template is model independent. The user is required to define a template model, while the code will efficiently search for the matching subjet-like structures in the jet energy flow patterns. 
 \item The pattern recognition based approach permits an efficient way of determining the jet topology while at the same time taking into account the event kinematics (jet mass, subjet asymmetry, etc.). 
 \item \template tools can be used to preserve as much energy flow information as possible, which is particularly useful for events where the energy distribution is all that is available. This information is presented in a well-organized form convenient for a detailed analysis of jet substructure. 
 \item Peak templates have well defined jet shapes insensitive or weakly sensitive to pileup and underlying event.

\end{itemize}

\template is a C++ library which provides basic implementation of the Template Overlap Method for jet substructure. We designed the code around the {\sc FastJet} \cite{Cacciari:2008gp} package of jet algorithms with the aim at easy implementation into existing jet analysis tools. 
\template is also a testbed containing programs and routines for generating template data sets and collecting and analyzing statistics on the performance of Template Overlap and jet shapes. The package can be downloaded from \url{http://tom.hepforge.org/}.

In addition to overlap analysis, \template provides the necessary tools to analyze a boosted jet using observables constructed out of
best matched templates. We provide implementations of various template-based jet shapes and energy flow observables such as Template Planar Flow, Template Angularity, etc.

This manual  describes how to download and install \template, how to use the main libraries, as well as how to change its configuration for different overlap measures and template catalogs. Finally, the manual shows how to use the sample programs for testing purposes and basic data analysis. In section 2, we briefly introduce the physics behind the commands of \template. We give a short description of the \template code structure in section 3. Section 4 lists several possibilities for further extensions of the program. Appendices A and B discuss boost-invariant implementations of the template generation and various jet shape observables. Appendix C contains the detailed syntax and functionality of all relevant internal methods.

\section{Physics Overview} \label{Sec:overlap}

The current version of \template allows the user to study 
the substructure of massive high-$p_T$ jets for various models. The user defines a model by specifying a
catalog of partonic decay configurations, labelled as $f$,  which are taken to represent the decays of a heavy particle of mass $M$ at a given $p_T$. In addition, 
one has to specify a functional measure to quantify agreement between the energy flow of a jet and the flow of each template.  For each jet candidate, the overlap function is defined as
\begin{equation}
 Ov_N = \max_{\{f\}}\, \exp\left[ -\sum_{a=1}^N \frac{1}{\sigma_a^2}\left( \epsilon \, p_{T,a} -\sum_{i\in j} p_{T,i} \,F(\hat n_i,\hat n_a^{(f)}) \right)^2 \right] \label{eq:Ov}
\end{equation}
where $\{f\}$ is the set of templates defined for the given jet $p_T$, $p_{T,a}$ are the transverse momenta of the heavy particle (or resonance) decay daughters for the given template, $p_{T,i}$ is the $p_T$ of the $i^{th}$ jet constituent (or calorimeter tower, topocluster, etc.). The parameter $\epsilon$ serves to correct for the energy emitted outside the template subcone.
The first sum is over the $N$ partons in the template and the second sum is over jet constituents. The kernel functions $F(\hat n, \hat n_a^{(f)})$ restrict the angular sums to (nonintersecting) regions surrounding each of the template momenta.  We provide two concrete implementations of kernels: a Gaussian around each of the directions of the template momenta with normalization $F(0, \hat n _a^{(f)}) = 1$,
\begin{equation}
F(\hat n_i,\hat n_a^{(f)}) = \exp\left[- (\Delta R)^2/(2\omega_a^2)\right],
\end{equation}
and a normalized step function that is nonzero only in definite, non-overlapping, angular regions around the directions of the template momenta $p_i$,
\begin{equation}
 F(\hat n_i,\hat n_a^{(f)}) = \left\{ \begin{array}{rl}
 1 &\mbox{ if $ \Delta R <R_a$} \\
  0 &\mbox{ otherwise}
       \end{array} \right. ,
\end{equation}
 where $\Delta R$ is the plain distance in the ($\eta$,$\phi$) plane.
The parameters $\omega_a$ ($R_a$) determine the radial scale of the template subjet. Together with the energy resolution scale $\sigma_a$, these are the only tunable parameters of the model.
A few possible strategies to determine the optimal values of  $\sigma_a$ and $R_a$ are as follows:

\begin{itemize}
 \item Choose the best parameters according to the  tagging efficiency and background rejection (e.g. fix the efficiency and vary the sub cone size to achieve best background rejection power). Use the same parameters in every event. 
 \item For each event, make a choice of parameters for which the overlap is maximized. Estimate the stability of the configuration. 
 \item Choose the parameters separately for each template, {\it e.g.} using a $p_T$-dependent scale for template matching. 
\end{itemize}

Template overlap provides a mapping of final states $j$ to partonic configurations $f[j]$ at any given order.
The best matched template $f[j]$ can be used to characterize the energy
flow of the jet, giving additional information on the likelihood that
the event is signal or background. Furthermore, we can derive additional jet shape information out of $f[j]$ to further increase the rejection power of the method. 

It is important to realize that other functional measures and kernel functions can easily be implemented, and we encourage the reader to explore them.
The choice of template parameters, kernels and functional measures is largely dependent on the application of Template Overlap and has to be determined on a case by case basis. The same is true for the template generation. For instance, a boosted Higgs analysis benefits from the use of both the two-body and three-body overlap analysis, whereas two-body overlap could be of little use in a boosted top analysis. 


\section{Template Overlap in a Pileup Environment} \label{sec:pileup}

\subsection{Effects of Pileup on Peak Overlap}
The high luminosity environment characteristic of the LHC provides an obstacle for all jet observables, with the recently finished $\sqrt{s} = 8$ TeV run resulting in stunning $\langle N_{vtx}\rangle = 20$ interactions per bunch crossing. Jet mass and transverse momentum as well as most jet substructure observables (see Ref. \cite{Soyez:2012hv} for instance) can be severely affected by pileup, prompting a need for various pileup subtraction and correction techniques. 

One of the valuable features of the Template Overlap Method is that it is weakly sensitive to pileup, even in a high pileup environment. There are three main reasons for this: 

\begin{enumerate}
\item Template Overlap Method is designed to detect large ``spikes'' in energy deposition, characteristic of the hard radiation. 
\item The peak overlap is sensitive only to the energy flow deposited within a template sub cone of radius $R_a$. To first order approximation, the effects of pileup on the value of overlap should be proportional to the area of the template sub-cones \footnote{For instance, consider a boosted Higgs jet clustered with a radius $R=1.0$ and analyzed with a two body template of radius $R_a = 0.1$. The relative pileup contribution to the overlap compared to the jet $p_T$ will be of the order $O(R_a^2 / R^2) = 0.01$.  }.
\item The typically small template sub-cone (relative to the jet cone size) protects the Template Overlap method from a large pileup contamination.
\end{enumerate}

\begin{figure}[htp]
 \centering
 \begin{center}
 \includegraphics[width=5in]{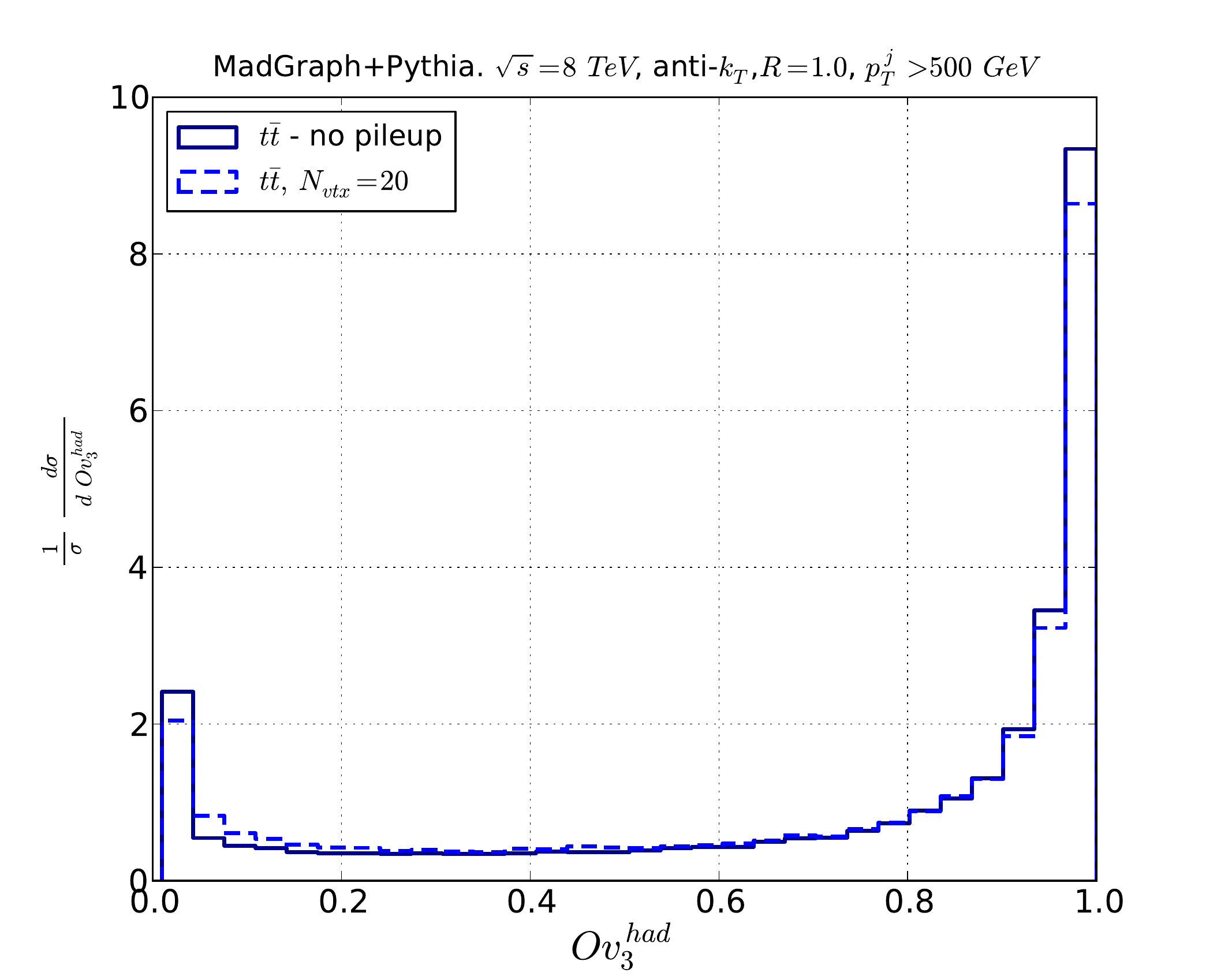}
\end{center}
\caption{ Effects of pileup on a three body peak overlap distribution of boosted top jets. The solid line shows events with no pileup, while the dashed line shows events $\langle N_{vtx}\rangle = 20$ interactions per bunch crossing.   \label{fig:temp_pileup}}
\end{figure}

Figure \ref{fig:temp_pileup} shows an example of the effect of $\langle N_{vtx} \rangle= 20$ pileup interactions on the distributions of three body peak overlap of boosted top jets. The distributions remain largely unaffected with ramifications of pileup resulting in $10 \%$ effects \cite{Backovic:2012jj}. The same is true for other template observables \cite{Backovic:2012jj}.

\subsection{Using Multiple Template $p_T$ bins in a High Pileup Environment}

When running the overlap analysis, it is often beneficial to use several sets of template libraries, each one with a different template $p_T$ in order to better cover the $p_T$ range of the signal region. In a no-pileup environment, selecting the appropriate template $p_T$ bin is trivial, whereby the transverse momentum of the fat jet determines the template appropriate template $p_T$.  Such a method is not always fitting in the presence of pileup as the $p_T$ of the fat jet is shifted to higher values, thus not accurately representing the $p_T$ of the underlying heavy particle of interest. There are two general ways to work around this issue without requiring pileup subtraction/correction methods:
\begin{enumerate}
\item Analyze each jet with every template library. The overlap analysis, by design, attempts to match the kinematics of the template to the kinematics of the hard part of the fat jet. Thus, the template library which gives the highest overlap will be the one with the $p_T$ closest to the transverse momentum of the hard part of the fat jet. This method is useful when analyzing events in which only jets are present. For instance, boosted $t\bar{t}$ events in which both tops decay hadronically. Notice however that this method also increases the computing time by a factor of roughly the number of template $p_T$ bins.
\item Use an observable which is less sensitive to pileup and is correlated with the fat-jet $p_T$. For instance, consider $Zh$ events in which $Z$ decays leptonically and Higgs decays to $b\bar{b}$. In the transverse plane, the Higgs and the $Z$ are emitted back to back, yielding $p_T^{l_1} + p_T^{l_2} \approx p_T^{h}, $ but with the major difference that the $p_T$ of the leptons is almost unaffected by pileup. Thus, one can use the transverse momentum of the $Z$ to determine which template $p_T$ bin to use for the overlap analysis. Ref. \cite{Backovic:2012jj} employs a similar method for the analysis of $Wh$ events with large missing $E_T$.

\end{enumerate} 

\section{Program Structure and Use}
We proceed to discuss the installation of  \template and execution of the example code. We also discuss the general structure of the code, for the benefit of the user who might wish to read or modify the source code.

\subsection{Installation}
\template runs on any architecture with a modern C++ compiler such as \ttt{g++} and an installation of  {\sc FastJet  \cite{Cacciari:2008gp} }. For the convenience of Unix and Mac OS X users, we provide Makefile scripts.  \template depends on {\sc FastJet} for finding jets and uses the internal {\sc FastJet} classes for implementation of basic relativistic kinematics. The current version of \template requires {\sc FastJet} version 3.0 or higher. To use the Makefile provided with the code, simply add the location where {\sc FastJet} is installed, so \ttt{fastjet-config} can be found.

To install and run the \template follow these steps  \footnote{The following instructions are only for a Unix based operating system such as Linux or Mac OS X.}:
\begin{enumerate}
\item In a web browser, navigate to 

 http://www.hepforge.org/archive/tom/
 \item  Download the (current) source tar-ball and extract it by typing:
 \begin{verbatim}
 tar -xvf TemplateOverlap-X.Y.Z.tar.gz
\end{verbatim}
and replacing \ttt{X.Y.Z} with the appropriate version number (currently 1.0.0).
This will create a new subdirectory \ttt{TemplateOverlap-X.Y.Z} where all the \template source files are now ready and unpacked.
\item Move to the resulting directory (\ttt{cd TemplateOverlap-X.Y.Z}) and compile one of the examples
\begin{verbatim}
cd TemplateOverlap-X.Y.Z/
g++ -Wall -O2 example.cc  -o example \
                   `${FASTJETLOCATION}/fastjet-config --cxxflags --libs`
\end{verbatim}
The code can also be compiled with the provided \ttt{Makefile} in environments where \ttt{make} is available.

\item The previous step compiles the \ttt{example} program which illustrates the basic functionality of the \template. The program reads a test event file \ttt{jet.dat} and a template file \ttt{template2b.dat} . To execute the test program, type
\begin{verbatim}
./example jet.dat template2b.dat
\end{verbatim}
The example code will write output to the terminal, \ttt{stdout},  and should read:
\begin{verbatim}
Hardest jet: pt, y, phi =  324.9   0   0, mass = 125.788
The best-matched templates are: (Ov2 = 0.968626)
pt, y, phi =  220.425 0.259875 0.0501516, mass = 0.
pt, y, phi =  105.432 -0.525 6.17819, mass = 0.
\end{verbatim}

\end{enumerate}

Now that you have successfully compiled and ran the \ttt{example} application (and perhaps even compiled and run it), we can proceed with a more detailed discussion of the \template algorithm and structure. The following explanation will provide you with a basic understanding of the functionality of the underlying \template code.

\subsection{The Algorithm}

The core of the \template package is a numerical implementation of the template matching algorithm of Eq. \ref{eq:Ov}. The two primary components of every template matching process are:
 \begin{itemize}
  \item {\bf Source event ({$j$}):} A jet containing full information about the constituents or calorimeter energy deposition. 
  \item{\bf Template ($f$):} A signal template which serves to construct a comparative measure.  
 \end{itemize}
The goal of Template Overlap analysis is to identify events $j$ with a good to templates $f$. Events with high match have a higher likelihood of being signal. The \template package performs the analysis in a sequence general enough to be applicable in a wide variety of HEP analyses:
\begin{enumerate}
 \item Generate sets (or ``catalogs'') of large number of $N$-body templates which uniformly cover the phase space of a massive particle decay of mass $M$ at a given $p_T$. We suggest that templates be generated in the lab frame by solving all the available kinematical constraints, as in Appendix~\ref{Sec:Templates}. For a realistic analysis, it is usually necessary to generate several sets of templates for different values of $p_T$ and dynamically determine which template set is appropriate based on jet $p_T$. \footnote{Choosing a template set based on jet $p_T$ could be inappropriate in a high pileup environment. See Section \ref{sec:pileup} for more detail.} 
 Alternatively, templates can be generated in the rest frame of the event and boosted to the lab frame on an event-by-event basis. Note however, that this method comes with a significant increase in computation load, as millions of four momenta will typically have to be boosted for each event. 
 
 \item For each template, calculate a measure $d(j,f)$ to quantify the match (how similar the energy flow of the template is to that particular region in the flow of the observed event) of the template and the event
\begin{equation}
d(j,f) = \,\exp\left[\, -\ \sum_{a=1}^N \frac{1}{2\sigma^2_a}\left(  \sum_{i\in \Omega}
   E_i^{(j)} F_N( \Omega,r)  - E_a^{(f)}\, \right)^2
   \right] ,
\end{equation}
 where $F_N(\Omega,r)$ is a kernel function and $r$ is the resolution scale parameter which determines the width and the shape of the kernel.  
 
 \item  For each template $f$ and event $j$, store the measure $d(f,j)$ in the result matrix $\bf  R$. The result matrix is analogous to the output of many image pattern recognition algorithms. Fig. \ref{Fig:color_map} shows an example. The points represent a complete template set at a fixed $p_T$ and $M$ of a two body boosted Higgs decay. The color map represents the value of $d(f,j)$  for each template state. The regions of high $d(f,j)$ are where most of the event $p_T$ was deposited.
 
\begin{figure}[htp]
 \centering
 \begin{center}
 \includegraphics[width=5in]{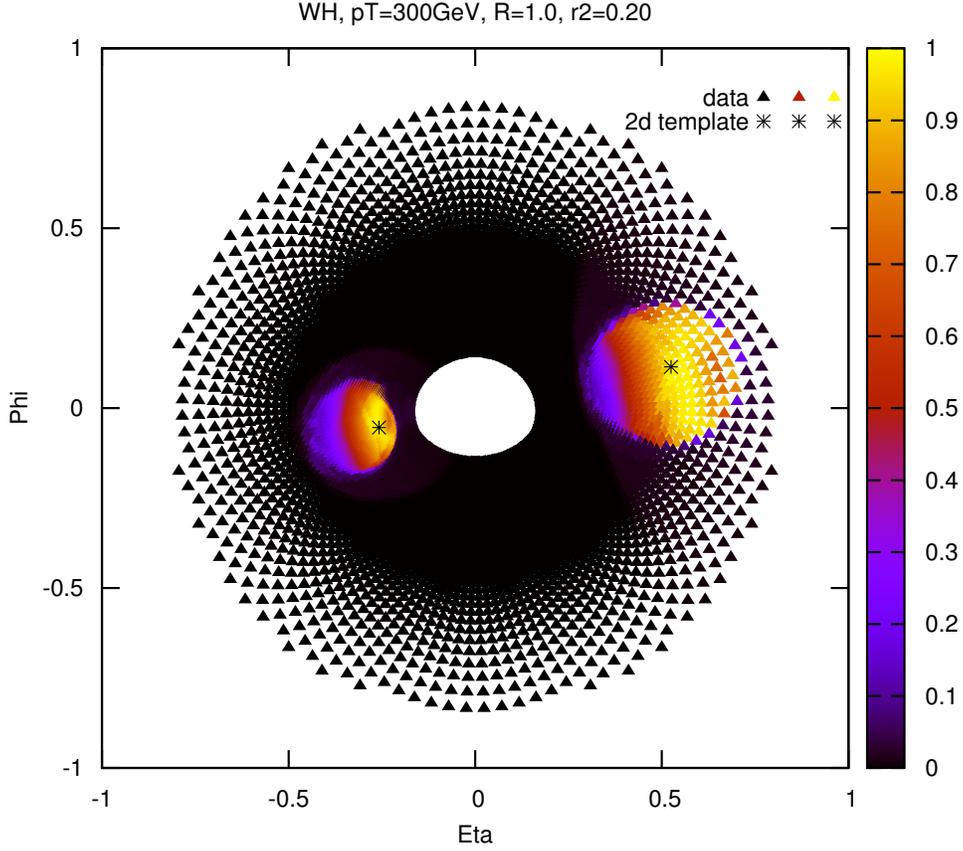}
\end{center}
\caption{Energy flow reconstruction from 2-particle templates for a single boosted Higgs event. The points show angular positions of  a highest $p_T$ template parton (two-particle templates). Note that the other parton is uniquely determined by energy conservation. The color map shows the overlap score of the template parton at various positions in $(\eta , \phi)$. The region around $\eta = \phi = 0$ is not covered due to the kinematic constraint of $\Delta R_{b\bar b} > 2m_h / p_{T}^h$. }\label{Fig:color_map}
\end{figure}

 \item For every event $j$, find the maximum value
 \begin{equation}
  Ov_{N} =  
{ \rm{max}}_{ \{f\}}\, d(j,f),
 \end{equation}
 where ${f}$ refers to maximizing over the entire set of templates. We refer to $Ov_N$ as the ``peak overlap.'' Similarly, we refer to the template $f_{max}$ which maximizes $d(f,j)$ as the peak template.

 \item The previous three steps are repeated as many times as necessary using different values of $N$, {\it e.g.} for a boosted Higgs, $N=2,3$.
 Fig.~\ref{Fig:event_view} shows a typical matching process for a Higgs jet with  $p_T=300\GeV$ analyzed with both 2- and 3-particle templates.
 \begin{figure}[htb]
 \centering
 \begin{center}
 \includegraphics[width=5in]{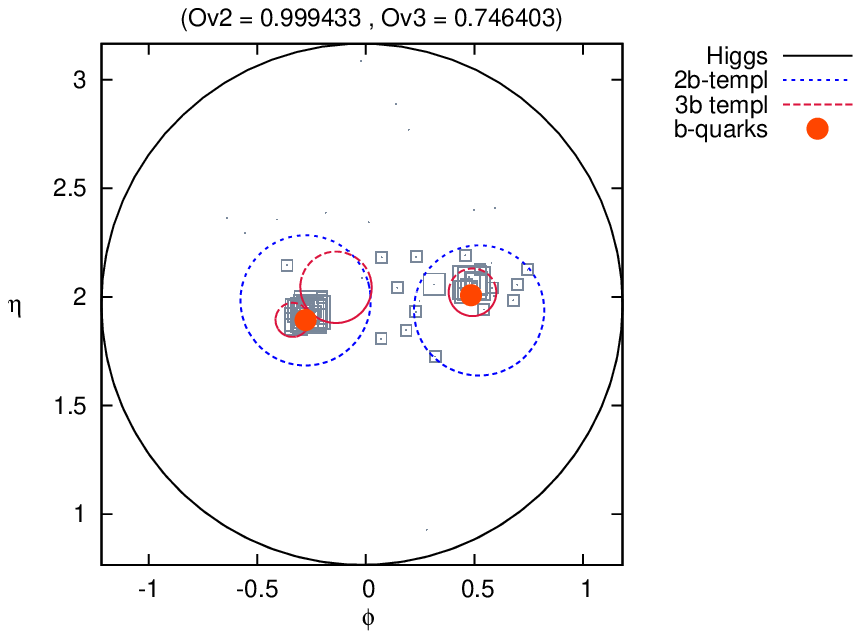}
\end{center}
\caption{Event displays for a typical Higgs jet with invariant mass near 125\GeV. The
blue and red circles indicate the region spanned by the best matched templates with $N=2,\,3$, respectively, using \ttt{CONE} and \ttt{DefaultMeasure}. 
In this and subsequent event displays, the particles are shown in grey cells of variable size, and the marker area for each cell is proportional its scalar transverse momentum. The solid red dots are the positions of $b$-quarks in the hard process.}\label{Fig:event_view}
\end{figure}

\end{enumerate}

\subsection{Efficient Generation of Template Libraries}

The \template package provides routines which generate template libraries for a given model. The $N$-particle template libraries in the $\eta-\phi$ space can be represented using $3N$-dimensional tables with equidistant grids in the $\eta,$ $\phi,$ and $p_T$ variables. The construction of such tables would typically proceed with the help of Monte Carlo data to determine the size of $p_T$ steps and the minimum number of templates  required to maximize signal efficiency while maintaining sufficient background rejection power. 

The function \ttt{TemplateBuilder} in \verb|build_template.cc| contains the full specification of how to carry out the generation of templates in the lab frame. According to the method described in Appendix~\ref{Sec:Templates}, a ``template definition'' should include parameters such as template $p_T$, mass and jet cone radius $R$ as well as the number of template patrons $N$. The function call is
\begin{lstlisting}
 void TemplateBuilder (std::ofstream & File,    
		      const fastjet::PseudoJet & axis,  
		      const double etaMax,  
		      const double phiMax,  
		      const double minPt, 
		      const int nEta, 
		      const int nPhi,
		      const int nPt, 
		      const double R,  
		      const int mode),
\end{lstlisting}
where \ttt{File} is the output file into which the code writes the template catalog and \ttt{axis} is the four vector of the event jet axis,  providing relevant parameters like template mass and $p_T$. \ttt{etaMax} and \ttt{phiMax} are the maximum value of $\eta$ and $\phi$ of a template parton relative to the jet axis, while \ttt{minPt} is the minimum $p_T$ of a template parton (for infra-red safety). \ttt{nEta}, \ttt{nPhi} and \ttt{nPt} are the number of steps in $\eta$, $\phi$ and $p_T$ respectively for the template generation. \ttt{R} is the anti-$k_T$ jet cone used to cluster the template patrons. If the patrons can not be clustered into a jet of radius \ttt{R}, the template is rejected. Finally, \ttt{mode} is one of the entries of the enumerated \ttt{TemplateModel}:
\begin{lstlisting}
enum TemplateModel {TOP, HIGGS2, HIGGS3, ...};
\end{lstlisting}
The current implementation of \template contains three default modes:
\begin{itemize}
 \item \ttt{TOP:} Three body $t$ decay.  The top template model generates template states to cover three-particle phase space for top decay, $t\to b + W\to b+ q+\bar q$, with the constraint $(p_q+p_{\bar q})^2 = M_W^2$. To construct these states, the algorithm uses a sequential scan over four angles. We take these to be the rapidities and azimuthal angles that define the $b$ and a $W$'s daughter in the lab frame, defined relative to the direction of the top jet axis.
 \item \ttt{HIGGS2:} Two-body Higgs decay. Two angles define the two-body state of the daughter particles. We choose these to be the rapidity and azimuthal angle of the first daughter in the lab frame defined relative to the  Higgs direction. 
 \item \ttt{HIGGS3:}  Three-body Higgs decay.  Four angles and one energy define the three-body state of the daugher particles. We take these to be the the rapidities and azimuthal angles that define the $b$ and a $\bar b$ directions in the lab frame, defined relative to the Higgs direction. The remaining variable is the $p_T$ of the leading parton. 
\end{itemize}

The \ttt{TemplateBuilder} routine surveys the kinematically-allowed template configurations by fixing the total four momentum to \ttt{axis} and then taking possible values of 
energies ($p_{T,i}$) and the angles ($\hat p_i$) within the bounded interval as defined by $\eta_{max}$, $\phi_{max}$ and $p_T^{min}$. The number of variables depends on the number of degrees of freedom of the configuration. The domain of the $3N-4$ independent variables, $\hat p_1, \cdots, \hat p_{N-1}$ and  $ p_{T,1},\cdots p_{T,N-2}$, that define a template
is divided into a uniform grid, according to a fixed interval, and the remaining transverse momentum $p_{T,N-1}$ is obtained by applying the restrictions of conservation of energy-momentum. 
The resulting groups which include negative $p_T$ are automatically discarded. An additional restriction of
\begin{equation}
 p_N = P -\sum_{i=1}^{N-1} p_i,
\end{equation}
imposes the condition $P = (p_T,0,0,E_J)$.

\subsection{Choosing the Kernel}
The optimal choice of the template matching kernel depends on the analysis strategy and the amount of information the user has about the signal and the background.  A reasonable choice of the kernel width typically assumes at least some kinematic properties or jet shapes of the signal. In fact, optimal choice will be different for different signals. For example, an analysis searching for a Higgs decaying into $b\bar b$ pairs is likely to make very different assumptions about jet substructure from a data analysis which looks for $t\bar t$ events in the all-hadronic 6-jet mode. It is thus both interesting and important to look at the substructure of a jet using a variety of kernels and kernel parameters. Table~\ref{Table:schemes} lists the kernels available in the default implementation of \template. 
 
The kernel function $F(\Omega, f)$ should be a sufficiently smooth function of the angles for any template state $f$ in order to ensure infra-red safety. For instance, the kernel could be defined as a Gaussian around each of the template momenta, which we provide with the option \ttt{GAUSSIAN}. Alternatively, we may choose $F$ to the a normalized step function that is nonzero only in definite angular regions around the directions of the template momenta $p_a$. This is the default option in \template and is implemented as an option \ttt{CONE}.

The \template package also allows the user to choose from a variety of template matching strategies to fix the energy resolution scales.  A fast and simple template matching can be performed using a single resolution scale (at one fixed cone radius or Gaussian width). This is the setting of the \ttt{FIXED} option and is the default option in \template.  Alternatively, a more sophisticated choice can also take into account more complex jet shapes. Indeed, the optimal width is not necessarily the same for every jet in an event, as low momentum subjets tend to have wider angular profiles. As an implementation of this feature, we propose varying cone schemes for template matching. The scheme is similar to the  fixed cone scheme, except that we allow for different cone radii to be associated to each template particle. 

Both the kernel and the energy resolution scale are set by variables in Table \ref{Table:schemes}.

\begin{table}[htb]
\centering
\begin{tabular}{|c|c|c|}
\hline
Enum & Default option& Alternate option\\
\hline
\ttt{Jet\_shape\_scheme} & \ttt{CONE}& \ttt{GAUSSIAN} \\ \hline
\ttt{Resolution\_scale\_scheme} & \ttt{FIXED} & \ttt{VARIABLE} \\

\hline
\end{tabular}
\caption{Members of the \ttt{Jet\_shape\_scheme} and \ttt{Resolution\_scale\_scheme} enums which define the choice of kernel function.}
\label{Table:schemes}
\end{table}

\subsection{Finding the Best Matched Templates}

\begin{figure}[!]
\begin{center}

\includegraphics[width=0.7\hsize]{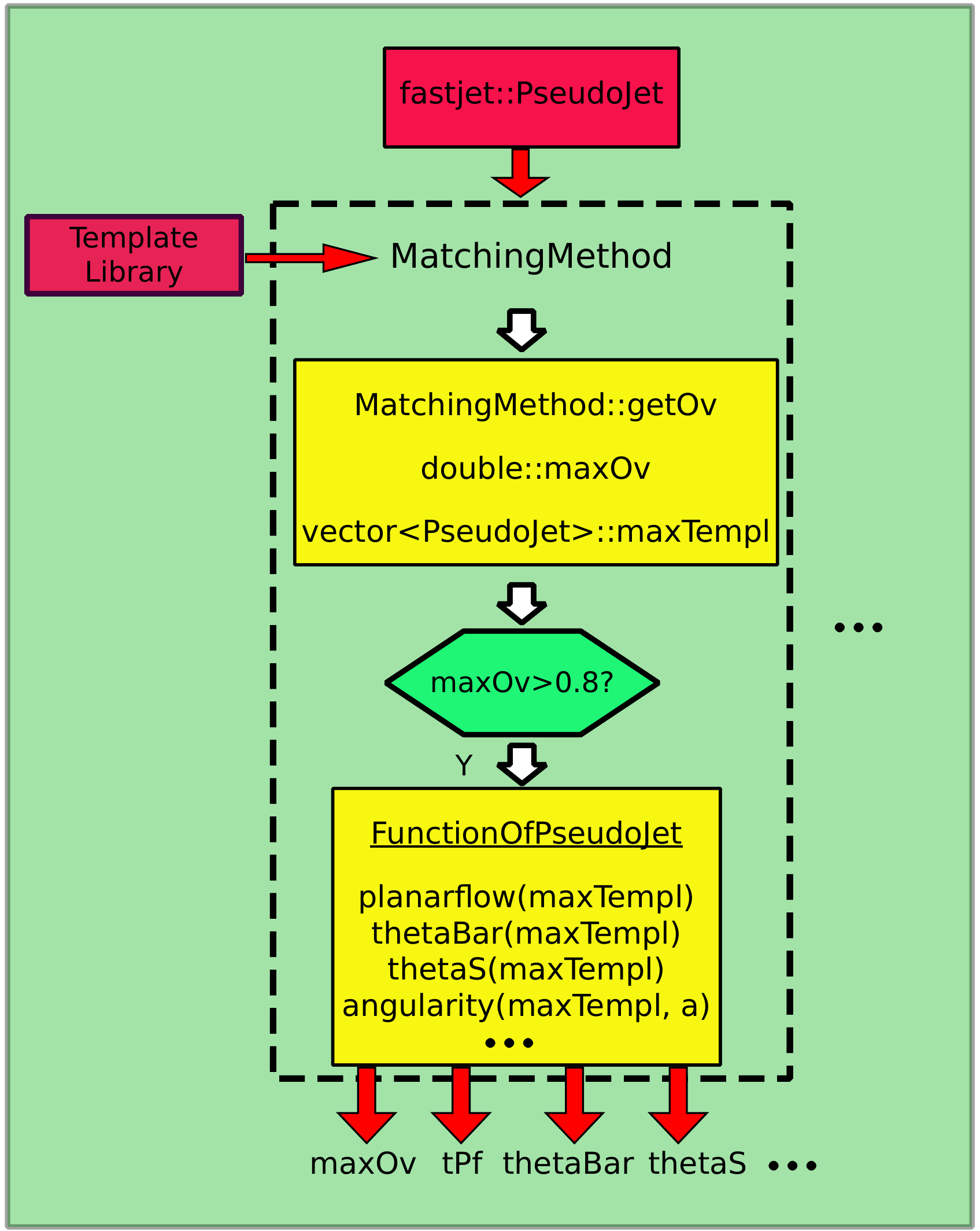}

\caption{
The structure of a typical \template analysis. A \ttt{MatchingMethod} gets input from a template catalog file using the default constructor \ttt{MatchingMethod::MatchingMethod}. The method \ttt{getOv} reads a \ttt{fastjet::PseudoJet} and passes the input through a sequence of template matching functions that are run sequentially. The output of \ttt{getOv} is then passed to several \ttt{FunctionOfPseudoJet}s that compute jet observables based on angular distributions of the partons in the best matched template.
}\label{Fig:flowchart}
\end{center}

\end{figure}

\ttt{MatchingMethod} class, the core of the \template package,  provides the implementation of the algorithm which is responsible for the Template Overlap analysis as a whole.  The algorithm is designed to perform template matching at fixed order in perturbation theory, with no constraint on the maximum level of complexity of energy flow patterns, {\it e.g.} when templates contain an arbitrarily large number of particles.   Most important part of the \ttt{MatchingMethod} declaration (in the ``matching.hh'' header file) is the constructor of the \ttt{MatchingMethod} class:

\begin{lstlisting}
	MatchingMethod(const string & templateFile, Settings mySettings)
\end{lstlisting}

The constructor arguments have the following meaning:
\begin{itemize}
 \item \ttt{templateFile}   A string containing the name of the template-catalog file.
 \item \ttt{mySettings} An object which collects several settings to use for template matching. See Section \ref{sec:settings} for more details.
\end{itemize}
 The simplest way of performing template matching with \template consists in constructing an object of this class at the beginning of the data analysis code and then running its \ttt{getOv()} member function on each jet.  Fig.~\ref{Fig:flowchart} shows a schematic representation of the algorithm.
 
The \ttt{MatchingMethod} loads the template catalog from a file using the default constructor. The member function \ttt{getOv} then reads a \ttt{fastjet::PseudoJet} and passes the input through a sequence of template matching functions which processes the following sequence:
\begin{itemize}
  \item Load an input event and a test configuration (template).
  \item Perform a template matching procedure by using the \ttt{TemplateOverlap} function \ttt{matchTemplate} with either of the two matching methods described before.
  \item Normalize the output of the matching procedure so that unity means perfect match. 
  \item Localize the template with the highest matching probability.
  \item Return the maximum value of overlap and best matched configuration (maximum overlap template).
 \end{itemize}
The \ttt{getOv} function returns the results of the maximization procedure in the format of \verb|temple_t| (defined in Section \ref{sec:misc}),
where the first element of the ordered pair is the value of the overlap and the second one contains a vector of template four momenta.
The user can pass the result of  \ttt{getOv} to several \ttt{FunctionOfPseudoJets} which compute jet observables from the momenta in the best matched template.

\newpage

\subsection{Sample Program} \label{sec:SampleProg}

An \ttt{example} program in  \ttt{short.cc} provides the basic functionality of \template. More specifically, the example shows how the \ttt{MatchingMethod} class can be used to perform basic substructure analysis with the Template Overlap method. 

 \ttt{example} requires only one input file: the template catalog. For the purpose of the example, consider a catalog of two-body templates stored in a file \ttt{template2bf.dat}. The command line call for \ttt{example} is: \begin{verbatim}
	./example  template2bf.dat
\end{verbatim}

Below, we give a detailed description of the relevant code snippets.

\begin{enumerate}
\item Include the appropriate header files. The core functionalities of \template are contained in the file \ttt{matching.hh}:
\begin{lstlisting}
 #include "matching.hh" 
 #include "fastjet/FunctionOfPseudoJet.hh"
 using namespace TemplateOverlap;
\end{lstlisting}

 \item Declare run parameters, event and result containers, as well as the match method. Set the jet cone radius $R = 1.0$ and the template sub-cone radius $r = 0.2$ as well as $\sigma_a =  p_T^a / 3$, where $p_T^a$ is the transverse momentum of the $a^{th}$ template parton. All settings are put into the \ttt{Settings} object:
 \begin{lstlisting}
  /// Set Run Parameters 
  double R = 1.2; // anti-kt parameter
  double R2 = 0.40; // template subcone radius
  double sigma = 0.333; // template Gaussian width
  myParams = Settings(R2, sigma, R);
  \end{lstlisting}

\item Define the event to be analyzed. For the purpose of the example, we hard-coded an event.
\begin{lstlisting}
  /// An event with three-particles
  std::vector<fastjet::PseudoJet> particles;
  particles.push_back( fastjet::PseudoJet( 112.0, -19.8, -56.1, 126.9 ));
  particles.push_back( fastjet::PseudoJet( 110.6,  13.9,  25.3, 114.4 ));
  particles.push_back( fastjet::PseudoJet( 102.3,   5.9,  30.8, 107.1 ));
\end{lstlisting}
 
\item Define the template matching instance which uses the template catalog of \ttt{template2bf.txt} and the parameters stored in \ttt{myParams}:
\begin{lstlisting}
  /// Create an instance of MatchingMethod for the analysis.  
  /// The ctor also loads the templates
  MatchingMethod myCone(argv[1], myParams); 
 \end{lstlisting}
 
 \item Analyze the event. \ttt{temp.first} contains the value of maximum overlap. \ttt{temp.second} contains the best matching template. 
\begin{lstlisting} 
   /// Find the best matched template
  templ_t result = myCone.getOv(jets[0]);
  std::vector<fastjet::PseudoJet> maxTempl = result.second;
  double maxOv = result.first;
  \end{lstlisting}

 \end{enumerate}
 
 The \ttt{getOv} method performs the core functions of the Template Overlap Method. As such, we deem it important to provide a more detailed discussion of its structure. 

 \begin{enumerate}
 \item Let us check the \ttt{getOv} function. First define the matching method and make a copy of the source event:
 \begin{lstlisting}
  /// Defining TemplateOverlap parameters
  int match_method = CONE;
  
   /// Source jet to matrix
  jet_t jet_mat; 
  myJet.copyTo(jet_mat);
 \end{lstlisting} 
 The \ttt{jet\_t} typedef is a vector of particles, that collects basic kinematic information about the constituents in a jet, {\it i.e.} their $\eta,\,\phi$ and $p_T$.
  \item Next, create the matrix that will store the results for each template location:
 \begin{lstlisting}
 /// Create the result matrix
  vector<double>  result; 
 \end{lstlisting}
 \item Use the \ttt{TemplateOverlap} function \ttt{matchTemplate} to search for matches between a template and a input event:
 \begin{lstlisting}
  /// Do the matching and normalize
  matchTemplate(jet_mat, _templates, result, match_method);
 \end{lstlisting}
 the arguments are naturally the input event $j$ (\verb|jet_mat|), the templates $f$ (\verb|_templates|), the result $\bf{R}$ (\ttt{result}), and the \ttt{match\_method}
 \item Use the \ttt{TemplateOverlap} function \ttt{maximize} to find the maximum overlaps (as well as their template directions) in the result array:
 \begin{lstlisting}
  /// Localizing the best match with minMaxLoc
  double maxVal; int maxLoc; 
  maximize(result, maxVal, maxLoc );
 \end{lstlisting}
The arguments of the function are:
 \begin{itemize}
  \item \ttt{result}: the source array.
  \item \ttt{maxVal}: variable to save the maximum value in the array, {\it i.e.} the maximum overlap.
  \item \ttt{maxLoc}: the point location of the maximum values in the array, {\it i.e.} the best match template.
 \end{itemize} 
 \item  Convert to \ttt{PseudoJet} and return the result:
 \begin{lstlisting}
  vector<PseudoJet> peak_template = ConvertToPseudoJet(_templates[_maxLoc], jet);  
  return std::make_pair<double, vector<PseudoJet> >(_maxVal, peak_template);
 \end{lstlisting}
\end{enumerate}
In summary, the output of \ttt{getOv} for any \ttt{PseudoJet} is the value of the overlap $Ov_{N}$, and also the identity of the peak template as a \ttt{vector} of \ttt{PseudoJet}s.

\section{Miscellaneous Tools} \label{sec:misc}

\subsection {Data types}

We have defined two new data types, \verb|templ_t| and \verb|jet_t|  for convenience. Here we briefly list the new data types for completeness and clarity, as they appear throughout the \template code. We typically use  \verb|templ_t| to store the results of the overlap analysis for each event. The \ttt{double} value typically holds the maximum overlap, while the vector of \ttt{PseudoJet} holds the best matching template. The definitions are:

\begin{enumerate}
\item
\begin{lstlisting}
 typedef std::pair<double, std::vector<fastjet::PseudoJet> > templ_t;
 \end{lstlisting}
 
\item
\begin{lstlisting}
 typedef std::vector<SingleParticle> jet_t;
\end{lstlisting}

\end{enumerate}

\subsection{\ttt{FastJet} plugin}

With the advent of {\sc FastJet 3.0+}, it has become straightforward to write wrappers for jet analysis tools around the suite of tools available in {\sc FastJet}. 
The {\sc FastJet} bare class \ttt{FunctionOfPseudoJet<T>} provides a common interface for jet measurements.  For the convenience of the user, we provide a class \ttt{Noverlap}, defined in \ttt{TemplateTagger.hh}, which performs all functions to the \template code within the {\sc FastJet} framework.  The class wraps the core Template Tagger code to provide the
 \ttt{ fastjet::FunctionOfPseudoJet} interface for convenience in larger analyses.  See \ttt{matching.hh} for definitions of $Ov_N$ and the constructor options.
The relevant methods of the class are:

\begin{enumerate}
\item 

\begin{lstlisting}
   Noverlap Noverlap(int N , double sigma, double R0, const string & templateFile)
\end{lstlisting}  

Constructor for the \ttt{Noverlap} class. \ttt{N} represents the order of the overlap analysis (two body, three body ...), \ttt{sigma} is the fraction of template parton $p_T^a$ used in $\sigma_a$. \ttt{R0} is the jet cone radius and \ttt{templateFile} is the file containing the template catalog. 

\item 
\begin{lstlisting}
  PseudoJet templ_t result(const PseudoJet& jet) const 
\end{lstlisting}

This function  returns the results of the overlap analysis in the format of a pair of values, the maximum value of overlap and the best matched template. See the discussion of \ttt{getOv} for more information. 

\end{enumerate}

{\sc FastJet 3.0+} also provides a common base class for jet manipulation: \ttt{Transformer}. Transformers can remove particles, re-arrange substructure, or tag/reject jets. The class \ttt{TemplateTagger} in \ttt{TemplateTagger.hh} implements a generic Template Overlap code described in the previous sections. 

The \ttt{TemplateTagger} class derives from \ttt{Transformer}, and can be constructed using a pointer to a \ttt{Selector} class derived from \ttt{FunctionOfPseudoJet<templ\_t>} which contains a value for the template kernel width, and the name of a file containing the catalog of templates. A simple example illustrates the implementation of \template within the {\sc FastJet} architecture: \begin{lstlisting}
  #include "TemplateTagger.hh"
  // ...
  /// Set up the template tagger
  SharedPtr<Noverlap> cone(new Noverlap(N, sigma, R2, argv[1]));
  SelectorMassRange(minMass,maxMass) selector;
  TemplateTagger tagger(cone.get(), selector, R2, ovcut);
  /// Now tag the leading jet using template tagger
  PseudoJet tagged_jet = tagger(jets);
\end{lstlisting} 

We adopt the convention that if a given jet does not satisfy  $Ov_N > $\ttt{ovcut}  the result of the transformer is a jet whose 4-momentum is zero. The \ttt{pieces()} of the resulting tagged jet correspond to the subjets that were associated to the best matched template:
\begin{lstlisting}
std::vector<fastjet::PseudoJet> subjets = tagged.pieces();
\end{lstlisting}
Additional structural information related to the value of the maximum overlap value and the best matched template is easily accessible. For instance:
\begin{lstlisting}
 cout << "(Ov2 = " << tagged.structure_of<fastjet::TemplateTagger>().ov() <<")"
       << endl << endl;
  cout << " The best-matched templates are: "<< std::endl;     
  PrintJets(tagged.structure_of<fastjet::TemplateTagger>().maxTempl()); ,
\end{lstlisting}
displays the value of maximum overlap and the four momenta of the best matched template. 

\subsection{Jet and Template Moments}
\template allows a user to calculate additional substructure observables. The {\sc FastJet 3} base class \ttt{fastjet::FunctionOfPseudoJet<double>} provides a 
common interface for the calculation. The following \ttt{FunctionOfPseudoJets} are available ( all defined in \ttt{TemplateTagger.hh}):

\begin{itemize}[label={}]
 \item \ttt{planarflow():} Calculates planar flow of the jet using longitudinally boost-invariant quantities.
 \item \ttt{angularity(int a):} Calculates the angularity $\tau_{-a}$ of the jet.
 \item \ttt{thetaS():} Calculates the angle between the softest template particle and the  jet axis. 
 \item \ttt{thetaBar():} Calculates an energy-unweighted distance between the template particles.
 \item \ttt{Stretch():} Calculates the imbalance in $\Delta R$ between the peak two-body template and the two leading subjets inside the jet.
 \item \ttt{Area():} Calculates the template area, {\it i.e.} the area in $\eta-\phi$ space projected by the cones around the directions of the template particles.
\end{itemize}

Each function returns a \ttt{double} value.  For a more detailed description of the above mentioned observables see Appendix \ref{sec:shapes}.

\subsection{\ttt{Settings}} \label{sec:settings}

The \ttt{Settings} class keeps track of all modes and parameters used during the jet clustering and template matching processes. As such, it serves all the \ttt{MatchingMethod} program elements from one central settings record. The user is allowed to access and change these settings to modify the template matching behavior. The complete list of methods and arguments is as follows.

\begin{lstlisting}
 /// Parameters that define TemplateOverlap
class Settings {
private:
   double _subConeRadius;     // subCone radius
   double _coneRadius;    // characteristic jet radius
   double _sigma;  // Gaussian energy resolution relative to parton pT
   double _coneRadius2;  // subjet radius
   int   _variableCone; // Dynamically change cone radius based on parton  pT (1 or 0)
   double _minPtParton; //For infrared safety, partons should not be to soft

public:
   Settings(const double subConeRadiusIn=0.20, 
	    const double sigmaIn=0.33, 
	    const double coneRadiusIn = 1.0,  
	    const double coneRadius2In=0.40, 
	    const int variableConeIn=1, 
	    const double minPtPartonIn = 10.) : 
   _subConeRadius(subConeRadiusIn), 
   _coneRadius(coneRadiusIn),
   _sigma(sigmaIn),
   _coneRadius2(coneRadius2In),
   _variableCone(variableConeIn),
   _minPtParton(minPtPartonIn) {}    //Default constructor.
   
   // Returns the value of the template sub cone radius
   double r() const {return _subConeRadius;}   
   //Returns the value for a jet cone radius 
   double R0() const {return _coneRadius;}   
   //Returns the radius of b-tagged subjets
   double R1() const {return _coneRadius2;}  
   //Returns the fraction of template parton p_T used as \sigma_a.
   double sigma() const {return _sigma;} 
   //Returns the status value for the overlap calculation mode, 
   //i.e. fixed or varying cone
   int variableCone() const {return _variableCone;} 
   //Returns the minimum p_T of the template parton with smaller transverse momentum. 
   double minPtParton() const {return _minPtParton;} 
};
\end{lstlisting}

\section*{Acknowledgments}
We thank Gilad Perez for introducing us to this topic and for a strong support in all stages of this project. We are very grateful to L. Levinson, P. Choukroun and the rest of the managers of Weizmann Institute physics computing farm for their flexibility and enormous help with large scale calculations. 
We thank S.J.~Lee and R.~Alon for the help with the early versions of the code and their suggestions and Gavin Salam for reading the code and sending valuable feedback. We also benefited greatly from discussions with P.~Sinervo J.~Winter,
F.~Spano, E. Duchovni and O.~Silbert. Finally, we would like to thank the CERN Theory group for their hospitality.

\appendix

\section{Properties of Templates }\label{Sec:Templates}

Template Overlap Method is a systematic framework 
aimed to identify kinematic characteristics of an boosted jet. A typical template configuration consists of a model template, $f$, calculated in perturbation theory, which describes a ``prong-like'' shape of the underlying hard subprocess of a jet. Template construction typically employs prior  theoretical knowledge of the signal kinematics and dynamics, as well as possible experimental input. For instance, Higgs 2-particle templates are sets of 2 four momenta which satisfy kinematic constraints of a boosted Higgs decay etc.

The simplest template configurations are the ones describing the kinematics of two-body processes such as the decay of SM Higgs or $W/Z$ bosons into quark-antiquark pairs. 
These are easily dealt with by assuming the rest frame of the parent particle
and producing two decay products with equal and opposite, isotropically-selected momenta and
magnitude, subject to energy conservation. The problem of a $N$-body decay subtracts four constraints from the decay products' $3 N$ degrees of freedom: three for overall conservation of momentum and one for energy\footnote{ For our purpose, we ignore the color and spin degrees of freedom.}. The final states can therefore be found on a $(3N-4)$-dimensional
manifold in the multi-particle phase space. Note that the dimensionality of the template space increases rapidly with additional patrons. For instance, the two body templates require only  two degrees of freedom, while a corresponding four body template space is already eight dimensional. 

The question of which kinematic frame the templates should be generated in requires careful consideration. 
Authors of Ref. \cite{Almeida2011aa} argued that a search for the global maximum of $Ov_N$
could be too computationally intensive. To improve the computation time, the template states were generated in the Higgs rest frame using 
a Monte Carlo routine, and then boosted into the lab frame. While this method worked sufficiently well for tagging
a highly-boosted object(\textit{i.e.} a $1\TeV$ Higgs jet), it introduced residual algorithmic dependence and a certain sense
of arbitrariness in the jet shape. At lower $p_T$ the Monte Carlo approach samples mainly the templates within the soft-collinear region, leaving other regions of phase space unpopulated. An enormous number of templates is required to adequately cover the phase space at $p_T \sim O(100 \GeV)$, thus fully diminishing the motivation for a Monte-Carlo approach. The simplest and most robust choice is then to generate templates directly in the lab frame and then rotate them into the frame of the jet axis. The result is a well covered template phase space in all relevant boosted frames. In addition, the lab frame templates result in a  significant decrease in  computation time as a much smaller number of templates are needed.

 We proceed to show how to generate the phase space 
for 2- and 3-parton final states as well as how to generalize the results to arbitrary $N$.

\subsection{The case of 2-body templates}

First, we summarize our notation and conventions. The model template consists of a set of four vectors, $p_1, \cdots, p_N$, on the hyperplane determined by the energy-momentum conservation,
\begin{equation}
 \sum_i p_i = P, \,\,P^2 =M^2, 
\end{equation}
where $M, P$ are the mass and four momentum of a heavy boosted particle, i.e. the Higgs.
For simplicity, we treat all template particles to be massless.
We work in an $(\eta,\, \phi,\, p_{T})$ space, where $\eta$ is pseudorapidity, $\phi$ azimuthal angle and $p_T$ transverse momentum. 
Without loss of generality, we can assume that the template points in the $x$ direction ($\eta=\phi=0$).  The templates are distributed according to 
\begin{equation}
p_i = p_{T,i}(\cos\phi_i,\sin\phi_i,\sinh\eta_i,\cosh\eta_i),\,\, i=1,2,3
\end{equation}
subject to the constraint
\begin{equation}
\sum_{i=1}^N p_i = P = (p_T,0,0,E_J) \label{momentum_cons}
\end{equation}
with $E_J=\sqrt{M^2+p_T^2}$. We find it useful to define unit vectors by
\begin{equation}
\hat p_i = (\cos\phi_i,\sin\phi_i,\sinh\eta_i,\cosh\eta_i),\,\, i=1,2, \label{unit_vector}
\end{equation}
so that $p_i = p_{T,i}\, \hat p_i$. 

Phase space for the 2-body decay processes is characterized by particularly simple kinematic parameters. To illustrate, 
first note that the 2-particle templates are uniquely determined by one single four momentum, $p_1$ subject to the condition
\begin{equation}
(P-p_1)^2=0.
\end{equation}
Writing $p_1 = p_{T,1} \hat p_1$, we can solve for $p_{T,1}$ in terms of the angles of the first parton
\begin{equation}
p_{T,1} = \frac{M^2}{2(P\cdot \hat p_1) }.
\end{equation}

We see that a 2-particle template is therefore completely determined in terms of the unit vector $\hat p_1$ as follows:
\begin{eqnarray}
 p_1 &= &\frac{M^2}{2(P \cdot \hat p_1)} \hat p_1 \\
 p_2 & = & P - p_1 .
\end{eqnarray}

Note that we can represent such a template as a point $(\hat \eta,\, \hat \phi)$  in $\eta-\phi$ plane. These are the two degrees of freedom, in accordance with the general result that the dimensionality of the $N$ template space is $3N-4$.

\subsection{The case of 3-body templates}

A space of five degrees of freedom allows for 3-particle templates to differ from one another in more than one way.
The 3-particle templates are determined by  two four momenta, $p_1$ and $p_2$, subject to the constraint, 
\begin{equation}
(P-p_1-p_2)^2=0.
\end{equation}

Using $p_1 = p_{T,1} \hat p_1$ and $p_2 = p_{T,2} \hat p_2$, we can solve for $p_{T,2}$ in terms of the angles of first two partons and $p_{T,1}$,

\begin{equation}
p_{T,2}= \frac{M^2-2 P\cdot p_1}{2( P\cdot \hat p_2 - p_1 \cdot \hat p_2 )}.
\end{equation}
A general 3-particle template is then completely specified by $p_{T,1}$ and two unit vectors (or, equivalently, four angles) $\hat p_1$ and $\hat p_2$.

\subsection{Extension to arbitrary $N$}

A generalization to an arbitrary number of particles is straight-forward. Proceeding as above, 
the $N$-particle templates are determined by $p_1,\cdots ,p_{N-1}$ subject to the constraint, 
\begin{equation}
(P-\sum_{i=1}^{N-1} p_i)^2=0.
\end{equation}
Using $p_i = p_{T,i} \hat p_i$, we can now solve for  $p_{T, N-1}$ in terms of the $p_1,\cdots ,p_{N-2}$ and $\hat p_{N-1}$,

\begin{equation}
p_{T,N-1} = \frac{M^2 + 2 \sum_{i<j}^{N-2} p_i\cdot p_j  - 2\, P\cdot \sum_{i}^{N-2} p_i }{2\, (\hat p_{N-1} \cdot P) - 2\,\hat p_{N-1}\cdot \sum_i^{N-2}p_i }
\end{equation}
For the special cases of $N=2$ and $N=3$, this formula reduces to the above results .

\section{Substructure and jet shapes} \label{sec:shapes} 

The \template code contains implementations of several jet shape observables in addition to the Template Overlap Method. 
Jet shapes are inclusive, infrared-safe observables which are smooth functions of the energy distribution within jets. They are constructed as
weighted sums over the four-momenta of the constituents of a
jet and reveal details about its inner structure, shedding light on its partonic origin.

\begin{itemize}
\item
A set of such jet shape observables is given by the class of angularities $\tau_a$ of a jet, defined by 
\begin{equation}\label{Eq:angularity}
 \tau_a \equiv \frac{1}{2 E_J} \sum_{i\in J} |{\bf p_T^i}| e^{-\eta_i(1-a)},
\end{equation}
where $a$ is a parameter taking values $-\infty < a < 2 $, the sum is over all the particles in the jet, $E_J$ is the jet energy, 
${\bf p}_T$ is the transverse momentum relative to the jet direction, and $\eta = - \ln \tan{\theta/2}$ is the pseudorapidity relative to the 
jet direction. 

Angularities, $\tau_a$,  are able to distinguish between QCD jets and other two-body decays.
Almeida {\it et al.} \cite{Almeida:2010pa} showed that the discriminating power of angularities is owned to the fact that the 
decays of color neutral objects are democratic, sharing energy symmetrically, whereas  
 QCD events with same mass are typically asymmetric. 

\item
Planar Flow ($Pf$)  \cite{Thaler:2008ju, Almeida:2008yp,GurAri:2011vx} is another useful jet substructure observable.
We defined the default \template implementation of $Pf$ in terms  of pseudorapidity $\eta = - \ln(\tan(\theta/2))$, the azimuthal angle $\phi$ and the transverse momentum $p_T$:
\begin{equation}
Pf= \frac{4\,{\rm det}\,{\bf I}}{({\rm tr}{\,\bf I})^2},
\label{Pfdef}
\end{equation}
where ${\bf I}$ is defined by,
\begin{equation}
{\bf I}= { \frac{1}{ m_J}} \sum_i {p_{T}^i}  \left( \begin{array}{ccc}
(\Delta \eta_i)^2 & \Delta\eta_i\Delta\phi_i  \\
\Delta\eta_i\Delta\phi_i  & (\Delta \phi_i)^2  \\
 \end{array} \right)
\end{equation}
with $m_J$ the jet mass, ${p_{T}^i}$ is the transverse energy of particle $i$ in the jet. Here, $(\Delta \eta_i, \Delta\phi_i) = \vec c_i-\vec J$, where $\vec J = (\eta_J , \phi_J )$ is
the location of the jet and $\vec c_i$ is the position of a cell or
particle with transverse momentum $p_{T}^i$. 
Notice that the $Pf$ definition of Eq. \ref{Pfdef} is invariant under boosts along the beam axis. 

Planar flow describes 
the way energy is deposited on the plane transverse to the jet axis. It peaks at zero for linear energy depositions
and is close to unity for uniform energy configurations.
For instance, two-pronged jets, such as leading order QCD jets, are expected to leave two cores of energy
 resulting in average low planar values of planar flow.  On the other hand,
 three-prong jets coming from hadronic
decays of boosted tops, are expected to have a rather uniform planar flow distribution. Thus
planar flow can be used to separate massive boosted QCD jets from top jets. 

\item Angular correlations of the template momenta which can otherwise be concealed in the numerical values of the peak overlap are of particular value. For instance, the angular distribution of a jet radiation can be measured with the variable $\bar \theta$, defined as
$$
	\overline{\theta} = \sum_i \sin \Delta R_{iJ},
$$
where $\Delta R_{iJ}$ is the distance in the $\eta-\phi$ plane between the $i^{th}$ template momentum and the jet axis. 
When measured using three-boy templates, the variable $\bar \theta$ characterizes the difference in angular ordering in our peak templates between the signal and background. Notice that for highly boosted jets, the 2-body version of $\overline{\theta}$ simply reduces to the angle between the two templates.

\item Template Stretch is a pileup insensitive observable sensitive to the mass difference between a jet and the peak template. First introduced in Ref. \cite{Backovic:2012jj}, template stretch is defined as: 
$$
S_{b\bar{b}} ^{(f)}= \frac{\Delta R_{b\bar{b}} }{ \Delta R_{f}}.  
$$
where $\Delta R_{f}$ is the distance between the peak two-body template momenta and $\Delta R_{b\bar{b}}$ is the distance between the two $b$-tagged sub-jets. A generalization of $S$ to non-$b$-tagged jets and other kinematic configurations is straightforward.

\end{itemize}

\section{\template Classes and Commands}

\subsection{Classes}

\subsubsection{\ttt{SingleParticle}}
\ttt{SingleParticle} is a helper class for \ttt{MatchingMethod} whose aim is to contain the minimum information about a particle in an event or a template, mainly its energy (or transverse momentum), rapidity and azimuthal angle. When the particle is associated with a template, there are two variables that contain additional, non-kinematics information: the template parton's width parameter \ttt{radius} (i.e. radius of the template sub cone), and its energy resolution \ttt{sigma}.
\begin{lstlisting}
 /// A helper class for MatchingMethod
class SingleParticle {

public:

  // Constructors.
  SingleParticle(  double pTIn = 0., double yIn = 0., double phiIn = 0.)
  :  pT(pTIn), y(yIn), phi(phiIn), mult(1), isUsed(false), radius(0.), sigma(0.) { }
  SingleParticle(const SingleParticle& ssj) :  pT(ssj.pT),
    y(ssj.y), phi(ssj.phi), mult(ssj.mult),
    isUsed(ssj.isUsed), radius(ssj.radius), sigma(ssj.sigma) { }
  SingleParticle& operator=(const SingleParticle& ssj) { if (this != &ssj)
    {  pT = ssj.pT; y = ssj.y; phi = ssj.phi; 
    mult = ssj.mult; isUsed = ssj.isUsed; 
    radius =ssj.radius; sigma=ssj.sigma;} return *this; }

  // Properties of particle.
  double pT, y, phi;
  int    mult; 
  bool   isUsed;
  double radius; //For templates
  double sigma;

  double deltaR2(const SingleParticle & other) const {
    double  dPhi = abs(phi - other.phi );
    if (dPhi > M_PI) dPhi = 2. * M_PI - dPhi;
    double  dEta =  y - other.y;
    return (dEta * dEta + dPhi * dPhi );
  }
  
};
\end{lstlisting}

\subsubsection{\ttt{MeasureFunctor}}
\ttt{TemplateTagger} provides a bare class to define custom overlap functions: \ttt{MeasureFunctor}. This provides the user with some flexibility to specify different kernel functions or to build new ones for customized template analyses. The \ttt{MeasureFunctor} provides the minimum bare class from which other overlap functions can be derived. The most important part of the class declaration looks as follows:

\begin{lstlisting}
class MeasureFunctor {
protected:
  MeasureFunctor() {}
public:
  virtual double distance(const SingleParticle& particle,const SingleParticle& axis)=0;
  virtual double numerator(const SingleParticle& particle,const SingleParticle& axis)=0;
  std::vector<double> subOverlaps(const jet_t & particles, const jet_t& axes);
  double overlap(const jet_t & particles, const jet_t& axes);
};
\end{lstlisting}

\subsubsection{\ttt{DefaultMeasure}}

The \ttt{DefaultMeasure} implements a cone-based template matching with the function
\begin{equation}
 F(\hat n_i,\hat n_a^{(f)}) = \left\{ \begin{array}{rl}
 1 &\mbox{ if $ \Delta R <R_a$} \\
  0 &\mbox{ otherwise}
       \end{array} \right. ,
\end{equation}
By default, this function finds all particles within a cone of radius $R$ from the template parton and calculates the unweighted sum of the particles $p_T$. By default, a fixed cone radius $R$ is used. This can be modified via the \ttt{Settings::set\_varying\_cone()}. This option corresponds to varying cone. Note that for the varying cone mode to work the user needs to specify a model for the cone scaling rule (energy profile). See \ttt{eShape} for more details.
\begin{lstlisting}
class DefaultMeasure : public MeasureFunctor {
private:
  Settings _settings;
public:
  DefaultMeasure(Settings settings): _settings(settings) {}
  virtual double distance(const SingleParticle& particle, const SingleParticle& axis){
         return std::sqrt(particle.deltaR2(axis)); }
  virtual double numerator(const SingleParticle& particle, const SingleParticle& axis){
         double deltaR = std::sqrt(particle.deltaR2(axis));
         if (deltaR > _settings.r()) return 0.0;
         return particle.pT;   }  };
\end{lstlisting}

\subsubsection{\ttt{GaussianMeasure}}

Similar to \ttt{DefaultMeasure}, but uses a Gaussian kernel function
$$
	F(\hat n_i,\hat n_a^{(f)}) = \exp\left[- (\Delta R)^2/(2\omega_a^2)\right],
$$
to add the $p_T$ of each particle in a jet. It has two constructors with the same arguments as \ttt{DefaultMeasure}.
\begin{lstlisting}
class GaussianMeasure : public MeasureFunctor {
private:
  Settings _settings;
public:
  GaussianMeasure(Settings settings): _settings(settings) {}
  
  virtual double distance(const SingleParticle& particle, const SingleParticle& axis) {
         return std::sqrt(particle.deltaR2(axis));}
  virtual double numerator(const SingleParticle& particle, const SingleParticle& axis) {
        double etaNow = particle.y;
	double phiNow = particle.phi;
	double rNow = _settings.r();
	double weight = exp(-(etaNow*etaNow+phiNow*phiNow)/(2 * rNow * rNow));
         return (particle.pT * weight);}
}; 
\end{lstlisting}

\subsection{Functions}

\subsubsection{\ttt{matchTemplate}}
The \ttt{matchTemplate} function is used to locate patterns inside the observed energy distributions within jets which have good overlap (``match'') to the set of templates. The algorithm can handle a variety of complicated patterns, {\it e.g.}, when the templates have more than the minimum number of partons or when there are additional kinematical constraints, such as the $W$ mass in a hadronically decaying top quark. The algorithm is implemented in the \ttt{matchTemplate} function whose prototype looks like this:
\begin{lstlisting}
void matchTemplate(jet_t & jet,
		   const  vector<jet_t> & templates, 
		   vector<double>  & result, 
		   int match_method);
\end{lstlisting}
The function arguments have the following meaning:
\begin{itemize}
 \item \ttt{jet:} The jet being analyzed. It must be a vector of \ttt{SingleParticle}.
 \item \ttt{templates:} Comparison template catalog. It must not have more particles than the source event.
 \item \ttt{result:} Map of comparison results. It must be a one-dimensional vector of \ttt{double}. After the search, its dimension is \ttt{templates.size()}.
 \item \ttt{match\_method:} Parameter specifying the functional measure or comparison method. It can take values available in \ttt{Jet\_shape\_scheme} . See Table \ref{Table:schemes} for details. 
\end{itemize}

\subsubsection{\ttt{maximize}}
Finds the global maximum in the result array and its position. 
\begin{lstlisting}
 void maximize(const vector<double>  & result, double & maxVal, int & maxLoc );
\end{lstlisting}

The parameters have the following meaning:
\begin{itemize}
 \item \ttt{result}: input array
 \item \ttt{maxVal}: pointer to the returned maximum value; \ttt{NULL} is used if not required. 
 \item \ttt{maxLoc}: pointer to the returned maximum location
\end{itemize}

\subsubsection{\ttt{ConvertToPseudoJet}}

For internal use. Converts a \ttt{jet\_t} to \ttt{PseudoJet}. 
\begin{lstlisting}
 std::vector<fastjet::PseudoJet> ConvertToPseudoJet(const jet_t& particles); 
\end{lstlisting}

\subsubsection{\ttt{ConvertToMat}}
For internal use. Converts a \ttt{PseudoJet} to \ttt{jet\_t}. 
\begin{lstlisting}
 jet_t ConvertToMat(const fastjet::PseudoJet& jet);
\end{lstlisting}

\subsubsection{\ttt{overlapDistance}}
Calculates the overlap between two jets or templates in the $\eta,\phi, p_T$ space, assuming one jet is a ``template''. \ttt{R0} is the template sub-cone radius. 
\begin{lstlisting}
 double overlapDistance (jet_t & jet1, jet_t & jet2, double R0) ;
\end{lstlisting}

\subsubsection{\ttt{reset}}
For internal use. Clears internal flags in a jet for reuse. 
\begin{lstlisting}
 void reset(jet_t & jet);
\end{lstlisting}

\subsubsection{\ttt{eShape}}
In some applications, one would like to have a more realistic model for the energy profile of a template parton than  simply a fixed cone. \template provides with a simple scaling rule for the template subcone radius that draws information from jet shape measurements at the LHC \cite{Aad:2011kq}. Note that the ATLAS jet shape study contains no data points for $p_T<30$ GeV. The \verb|eShape| extrapolates into the low $p_T$ region, with the constraint that the maximum value for the template radius is $0.3$. See Ref. \cite{Backovic:2012jj} for more detail.

To compute the energy profile of a template parton, the numerical values for the integrated jet shape measured by ATLAS are fit in different regions of jet $p_T$. \ttt{eShape} returns the radius, $r$, that is needed to contain 80\% of the transverse momentum in a cone of radius $r$ around the template parton direction.

\begin{lstlisting}

double eShape (double x)
{
  double aux =  0.422258 - 0.00377161* x + 0.0000174186 * x*x - 
  3.50639e-8 * x *x*x + 2.53302e-11 * x*x*x*x;
  
  return aux;
}
\end{lstlisting}

\bibliographystyle{unsrt}
\bibliography{template}

\end{document}